\documentclass{amsart}
\usepackage{amsmath,amssymb,euscript,graphicx,upgreek,stackrel}
\newcommand\cH{\EuScript{H}}
\newcommand\cJ{\EuScript{J}}
\newcommand\cO{\EuScript{O}}
\newcommand\cP{\EuScript{P}}

\newcommand\cV{\EuScript{V}}

\newcommand{\bC}{\mathbb{C}}
\newcommand{\bR}{\mathbb{R}}
\newcommand{\bZ}{\mathbb{Z}}
\newcommand{\bP}{\mathbb{P}}

\newcommand\ad{\operatorname{ad}}

\newcommand\End{\operatorname{End}}

\newcommand\tr{\operatorname{tr}}
\newcommand\str{\operatorname{str}}

\newcommand\A{\mathsf{A}}
\newcommand\C{\mathsf{C}}
\newcommand\F{\mathsf{F}}
\newcommand\Q{\mathsf{Q}}
\newcommand\U{\mathsf{U}}
\newcommand\si{\mathsf{i}}
\newcommand\sr{\mathsf{r}}
\DeclareMathOperator{\im}{im}
\DeclareMathOperator{\id}{id}

\DeclareMathOperator{\SO}{SO}
\DeclareMathOperator{\Or}{O}
\DeclareMathOperator{\Spin}{Spin}
\DeclareMathOperator{\rU}{U}
\newcommand\pdr\partial

\newcommand\al\alpha
\newcommand\gam\upgamma
\newcommand\del\delta
\newcommand\eps\epsilon
\newcommand\lam\lambda
\newcommand\sig\upsigma
\newcommand\tht\theta
\newcommand\vro\upvarrho
\newcommand\om\omega
\newcommand\upom\upomega
\newcommand\Gam\varGamma
\newcommand\Del\varDelta
\newcommand\Lam\varLambda
\newcommand\Om\varOmega

\newcommand\lpdr{\overset{{}_\leftarrow}{\pdr}}
\newcommand\rpdr{\overset{{}_\rightarrow}{\pdr}}
\newcommand\rnabla{\overset{{}_\rightarrow}{\nabla}}
\newcommand\hot{\hat\otimes}
\newcommand\extb{\ext\bullet}
\newcommand\Cl{\C_\hbar(V^*_\bC)}
\newcommand\pdm{\pdr_\mu}
\newcommand\pdn{\pdr_\nu}
\newcommand\ext[1]{\scalebox{1.1}[1.1]{$\mathsf{\Lambda}$}^{#1\,}}
\newcommand\sym[1]{\mathsf{S}\scalebox{1.1}[1.1]{${}^{#1\,}$}}
\newcommand\medext{\scalebox{1.1}[1.1]{$\mathsf{\Lambda}$}^2\,}

\newcommand\mfrac[2]{\raisebox{.5pt}
    {\scalebox{.85}[.85]{$\displaystyle\frac{\raisebox{-2pt}{${#1}$}}{{#2}}$}}}
\newcommand\mint{\raisebox{1pt}{\scalebox{.8}[.8]{$\displaystyle\int$}}}

\newcommand\mn{{\mu\nu}}

\newcommand\ii{\sqrt{-1}}
\newcommand\hf{\tfrac12}
\newcommand\qt{\tfrac14}
\newcommand\oP{\overline P}
\newcommand\PV{{\Pi V}}

\numberwithin{equation}{section}

\newcommand\bra\langle
\newcommand\ket\rangle

\begin{document}
\hfill {\tt arXiv:1909.nnnnn}[math-ph]\\

\begin{center}
{\bf\LARGE Representations of fermionic star product algebras\\
and the projectively flat connection}
\vspace{1.5em}

Siye Wu\vspace{1ex}

{\small Department of Mathematics, National Tsing Hua University,
Hsinchu, 30013, Taiwan\\
{\em E-mail address}: {\tt swu@math.nthu.edu.tw}}\\
\end{center}
\vspace{1em}

\begin{quote}
{\bf Abstract.}
We construct a family of fermionic star products generalising the fermionic
Moyal product.
The parameter space contains the polarisations necessary to define a quantum
Hilbert space.
We find a star product of fermionic functions on sections of the pre-quantum
line bundle and show that the star product of any function on a quantum
state remain a quantum state.
Associativity implies a representation of the fermionic star product
algebra on the quantum Hilbert space.
The star product is compatible with both the flat connection on the bundle
of fermionic functions and the projectively flat connection on the bundle of
Hilbert spaces over the space of polarisations.

\end{quote}
\vspace{1em}

\section{Introduction}
Given a classical system described by a symplectic or Poisson manifold,
deformation quantisation \cite{BFFLS77,BFFLS78} produces the quantum operator
algebra as a deformation of the algebra of functions in which the usual
product of two functions is replaced by a non-commutative but associative
star product, in a power series of $\hbar$, so that the first order term is
related to the Poisson bracket.
If $\hbar$ is treated as a formal parameter, existence of such deformation was
shown for symplectic manifolds in \cite{DL,OMY91,Fe} and for Poisson manifolds
in \cite{Kon}.
If $\hbar$ is a positive number, convergence of such formal series is known
only in limited cases, but most notably when the phase space is a symplectic
vector space.
The Moyal product \cite{Mo} of two functions on a symplectic vector space,
making use of the anti-symmetric Poisson bi-vector, corresponds to the Weyl
ordering in quantum mechanics, and it can be modified to fit various other
types of normal orderings in physics \cite{AW,Ba,Za,OMMY03}.
More generally, the Moyal product can be modified by a general symmetric
bi-vector, and various such star products are equivalent by explicitly
constructed intertwining operators \cite{OMMY05,OMMY07}.
Thus there is a family of star products on a symplectic vector space
parametrised by the symmetric bi-vectors, and the intertwining operators
define a flat connection over the parameter space.

On the other hand, geometric quantisation \cite{Ko70,So} is so far mainly
concerned with obtaining a quantum Hilbert space from classical data, which
include a phase space (symplectic manifold) with a pre-quantum line bundle
over it.
A quantum Hilbert space depends on additional data such as a polarisation,
which plays no role in classical physics.
In the general situation, there are few results on how the Hilbert space
depends on the choice of polarisations or how to best quantise functions as
operators on the Hilbert space if their Hamiltonian flows do not preserve
the polarisation.
However, if the phase space is a symplectic vector space and if the
polarisations are restricted to those given by compatible linear complex
structures, then there is a projectively flat connection on the bundle of
Hilbert spaces over the space of such complex structures \cite{ADPW},
providing a natural identification of quantum states in different
polarisations.
This projectively flat connection becomes flat upon metaplectic correction.
Moreover, the base space is identified as the Siegel upper-half space,
which is a non-compact Hermitian symmetric domain, and parallel transport
along the geodesics (with one or both ends at infinity) are the standard
Segal-Bargmann and Fourier transforms \cite{KW}
(see \cite{Wu11,Wu15,Wu17} for further development).

In \cite{WY}, a non-formal star product is defined of a function on a
symplectic vector space on a section of the pre-quantum line bundle over it,
for each compatible linear complex structure.
It has the important property that the star product of any function on a
holomorphic section remains a holomorphic section.
The star product is a deformation of the prequantum action in the sense that
the terms of $0$th and $1$st orders in $\hbar$ are precisely the prequantum
action of functions on sections.
A compatible complex structure determines a symmetric bi-vector on the
symplectic vector space and hence a star product of functions.
Another property of this star product is the associativity, making it a
representation of the star product algebra of functions on the space of
quantum states.
(See the references in \cite{WY} for other work on constructing such
representations.)
In addition, the star product is compatible with both the flat connection on
the bundle of functions and the projectively flat connection on the bundle of
Hilbert spaces \cite{WY}.
Using Fedosov's method \cite{Fe}, the construction in \cite{WY} is generalised
to a formal star product of functions on sections of the pre-quantum line
bundle on a general symplectic manifold \cite{Wu19}.

The fermionic phase space is a Euclidean vector space \cite{Ca,BM} and
the Poisson bi-vector is symmetric.
The classical observables form the exterior algebra of the phase space.
If the phase space is finite dimensional, deformation of the exterior
product to a fermionic star product that is not graded-commutative but
associative, written as a power series of $\hbar$, is necessarily non-formal.
The fermonic counterpart of the Moyal product was studied in
\cite{Bo,Za,HH,HHS04,GGPT}.
Following the geometric quantisation of bosons, pre-quantisation and
quantisation of fermionic systems were carried out in \cite{Ko77,Wo}.
To obtain the quantum Hilbert space, which is in fact the spinor
representation of the Clifford algebra, a polarisation from a complex structure
is again needed.
The allowed complex structures form a compact Hermitian symmetric space
and the spinor representations from different polarisations form a vector
bundle with a projectively flat connection, allowing identification of quantum
states from different polarisations by parallel transport \cite{Wu15}
(see also \cite{Wu11,Wu17}).
The projectively flat connection becomes flat upon fermionic metaplectic
correction \cite{Wu15}.

In this paper, we relate the approaches of deformation and geometric
quantisation of fermionic systems by carrying out a fermionic analogue of
\cite{WY}.
In \S\ref{sec:poisson}, we recall the geometry of fermionic phase space
and the Poisson bracket of functions of fermionic variables.
In \S\ref{sec:starK}, we generalise the fermionic Moyal product to a family
of star products parametrised by anti-symmetric bi-vectors (instead of
symmetric ones in the bosonic case).
There are intertwining operators relating different star products and they
define a flat connection on the bundle of exterior algebras over the parameter
space.
In \S\ref{sec:polar}, we study the geometry of the space of polarisations
and the fermionic star products associated to them.
In \S\ref{sec:qtn}, we recall the procedures of prequantisation and
quantisation of fermionic systems and the projectively flat connection on
the bundle of Hilbert spaces thus obtained.
In \S\ref{sec:starP}, we define the star product of a fermionic function on
a state for each polarisation and show that the result remain a state.
We show associativity and thus obtain a representation of the fermionic
star product algebra on the space of quantum states.
Furthermore, the star product is compatible with both the flat connection on
the bundle of exterior algebras and the projectively flat connection on the
bundle of quantum Hilbert spaces.
In \S\ref{sec:meta}, we define the star product of a function on a state
with metaplectic correction and demonstrate the compatibility with the flat
connection on the bundle of metaplectically corrected quantum states.
In the Appendix, we show that the star products of fermionic functions
actually comes from various quantisation maps from the exterior to the
Clifford algebra, giving alternative understanding of their properties.
We also express the fermionic star products of two fermionic functions as
fermionic integrals involving the two functions and an integral kernel.

\section{Fermionic phase space and the Poisson bracket}\label{sec:poisson}
The phase space of a linear fermionic system is a fermionic copy $\PV$ of a
Euclidean vector space $V$ in the sense that the complex fermionic `functions'
on it form the complexified exterior algebra $\extb V^*_\bC$ \cite{Ca,BM}.
Assume that the system has finite degrees of freedom so that $V$ is of real
dimension $m$.
The symmetric Euclidean structure $q\in\sym{2}V^*$ on $V$ replaces the
anti-symmetric symplectic form for a bosonic system.
The fermionic Poisson structure $q^\sharp\in\sym{2}V$, defined as the inverse
of $q$, is also symmetric.
We require that $\PV$ is oriented, with a positive unit volume element
$\eps_V\in\ext{m}V$.
The Berezin integral of a function $f\in\extb V^*_\bC$ on $\PV$ is
defined to be the pairing $\int_\PV f:=\bra f,\eps_V\ket$.

Let $\{e_\mu\}_{1\le\mu\le m}$ be a linear basis of $V$ with a positive
orientation and let $\{e^\mu\}_{1\le\mu\le m}$ be the dual basis of $V^*$.
Writing $q=q_\mn e^\mu\otimes e^\nu$, $q^\sharp=q^\mn e_\mu\otimes e_\nu$,
the components $q_\mn=q_{\nu\mu}$ and $q^\mn=q^{\nu\mu}$ satisfy
$q_{\mu\lam}q^{\lam\nu}=\del_\mu^\nu$.
The fermionic `coordinates' $\tht^\mu$ ($\mu=1,\dots,m$) on $\PV$ satisfy,
among other properties, the anti-commutativity law
$\tht^\mu\tht^\nu=-\tht^\nu\tht^\mu$.
A function on $\PV$ is of the form $f=\sum_{p=0}^m\mfrac1{p!}\,
f_{\mu_1\cdots\mu_p}\tht^{\mu_1}\cdots\tht^{\mu_p}$
and its Berezin integral can be written as
\[ \mint_{\!\!\PV}\,f=\mint_{\!\!\PV}\,d\tht\,f(\tht)=
\mfrac1{m!}\,\eps^{\mu_1\cdots\mu_m}f_{\mu_1\cdots\mu_m}, \]
where $\eps^{\mu_1\cdots\mu_m}:=\bra e^{\mu_1}\wedge\cdots\wedge e^{\mu_m},
\eps_V\ket$ is a totally anti-symmetric tensor.
If the basis $\{e_\mu\}$ of $V$ is chosen so that $\eps^{12\cdots m}=1$, then
we can write $d\tht=d\tht^m\cdots d\tht^2d\tht^1$ and the Berezin integral is
given by
\[ \mint_{\!\!\PV}\,d\tht^m\cdots d\tht^2d\tht^1\;
   \tht^{\mu_1}\tht^{\mu_2}\cdots\tht^{\mu_p}=0\quad(p<m),\qquad
\mint_{\!\!\PV}\,d\tht^m\cdots d\tht^2d\tht^1\;\tht^1\tht^2\cdots\tht^m=1. \]

The fermionic derivatives $\pdm=\mfrac\pdr{\pdr\tht^\mu}$ ($\mu=1,\dots,m$)
act as super-derivations on the space of `functions' on $\PV$.
Along a vector $v=v^\mu e_\mu\in V$, the directional derivative
$\pdr_v=v^\mu\pdr_\mu$ is also a super-derivation.
It will be useful to adjust the sign by redefining
$\pdr'_\mu f:=(-1)^{|f|-1}\pdm f$ or $\pdr'_vf:=(-1)^{|f|-1}\pdr_vf$ (for
$v\in V$), where $|f|$ is the degree of $f$ if it is homogeneous.
Then we have $\pdr'_\mu\tht^\nu=\pdm\tht^\nu=\del_\mu^\nu$.
The Poisson bracket of two fermionic functions $f,g$ on $\PV$ is
\[ \{f,g\}=\hf q^\mn\pdr'_\mu f\pdn g
 =-\hf f\big(q^\mn\lpdr_\mu\rpdr_\nu\big)g,  \]
where $\rpdr_\nu$ acts to the right on $g$ as usual while $\lpdr_\mu$ acts to
the left on $f$ so that $f\lpdr_\mu=(-1)^{|f|}\pdm f=-\pdr'_\mu f$; the sign
$(-1)^{|f|}$ is from exchanging the order of $\pdm$ and $f$ in the expression.
In particular, $\{1,f\}=0$ for all $f$ and $\{\tht^\mu,\tht^\nu\}=\hf q^\mn$
or $\{a,b\}=\hf q^\sharp(a,b)$ if $a,b\in V^*_\bC$ are regarded as (linear)
fermionic functions on $\PV$.
The Poisson bracket can be expressed as $\{f,g\}=\hf H_{\!f}\,g$ using the
fermionic analogue of the Hamiltonian vector field of $f$ on $\PV$,
\begin{equation}\label{eqn:ham}
H_f:=q^\mn\pdm'f\pdn.
\end{equation}

We can express the Poisson bracket as the restriction to the diagonal of a
fermionic function on $\PV\times\PV$.
Note that $\extb(V\times V)^*_\bC\cong\extb V^*_\bC\hot\,\extb V^*_\bC$,
where $\hot$ denotes graded tensor product.
The operator $q^\mn\pdm\hot\pdn$ acts by
$(q^\mn\pdm\hot\pdn)f\hot g=(-1)^{|f|}q^\mn(\pdm f)\hot(\pdn g)$,
where the sign comes from exchanging the order of $f$ and $\pdn$.
Let $\Del\colon V\to V\times V$ be the diagonal map $v\mapsto(v,v)$ and let
$\Del^*\colon\extb(V\times V)^*_\bC\to\extb V^*_\bC$ be its pull-back of
functions.
Then $\Del^*(f\hot g)=f\wedge g$ and the Leibniz rule
$\pdm(f\wedge g)=(\pdm f)\wedge g+(-1)^{|f|}f\wedge(\pdm g)$
becomes the operator identity
\begin{equation}\label{eqn:leib}
\pdm\circ\Del^*=\Del^*\circ(\pdm\hot1+1\hot\pdm)
\end{equation}
on $\extb(V\times V)^*_\bC$; note again the sign in
$(1\hot\pdm)f\hot g=(-1)^{|f|}f\hot(\pdm g)$.
The Poisson bracket of $f,g$ can be written as
\begin{equation}\label{eqn:poisson}
\{f,g\}=-\hf\Del^*((q^\mn\pdm\hot\pdn)f\hot g).
\end{equation}

Under the Poisson bracket, fermionic functions on $\PV$ form a Lie
superalgebra.
This can be verified easily by \eqref{eqn:poisson}.
For example, to show graded anti-commutativity, we use the graded flip
operator $\sig_2$ on $\extb V^*_\bC\hot\,\extb V^*_\bC$ defined by
$\sig_2(f\hot g)=(-1)^{|f||g|}g\hot f$, which satisfies
$\Del^*\circ\sig_2=\Del^*$ and $\sig_2\circ(\pdm\hot\pdn)
=-(\pdn\hot\pdm)\circ\sig_2$.
Therefore for all fermionic function $f$ and $g$ on $\PV$, we have
\[ \{f,g\}=-\hf\Del^*\sig_2((q^\mn\pdm\hot\pdn)f\hot g)
   =\hf\Del^*((q^\mn\pdn\hot\pdm)\sig_2(f\hot g))
   =-(-1)^{|f||g|}\{g,f\}.   \]
To show the graded Jacobi identity, we calculate, for all fermionic functions
$f,g,h\in\extb V^*_\bC$,
\begin{align*}
\{\{f,g\},h\}&=\qt\Del^*(q^\mn\pdm\hot\pdn)
   (\Del^*(q^{\lam\rho}\pdr_\lam\hot\pdr_\rho)f\hot g)\hot h   \\
&=\qt q^\mn q^{\lam\rho}\Del^*(\Del\times1)^*
(\pdm\pdr_\lam\hot\pdr_\rho-\pdr_\lam\hot\pdm\pdr_\rho)\hot\pdn(f\hot g\hot h).
\end{align*}
Here the Leibniz rule \eqref{eqn:leib} is used to obtain the second equality.
Note also that $q^\mn=q^{\nu\mu}$ and that
\begin{equation}\label{eqn:diag}
(\Del\times1)\circ\Del=(1\times\Del)\circ\Del=\Del_{(3)},
\end{equation}
where $\Del_{(3)}\colon V\to V\times V\times V$ is the tri-diagonal map
$v\mapsto(v,v,v)$.
It follows that
\begin{equation}\label{eqn:jacobi}
\{\{f,g\},h\}+(-1)^{|f|(|g|+|h|)}\{\{g,h\},f\}
+(-1)^{|g|(|f|+|h|)}\{\{h,f\},g\}=0.
\end{equation}
Finally, the fermionic Poisson bracket satisfies the graded Leibniz property
\[ \{f,g\wedge h\}=\{f,g\}\wedge h+(-1)^{|f||g|}g\wedge\{f,h\}. \]

The special orthogonal group $\SO(V,q)$ preserving the Euclidean structure
$q$ and the orientation of $V$is the group of linear canonical transformations
on the phase space $\PV$.
Consequently, it preserves the fermionic Poisson structure, i.e.,
$(\gam\otimes\gam)q^\sharp=q^\sharp$ for all $\gam\in\SO(V,q)$.
It acts as automorphisms on the exterior algebra $\extb V^*_\bC$ of fermionic
functions on $\PV$ by $f\mapsto\gam^\cO(f):=f\circ\gam^{-1}$.
The fermionic Poisson bracket is equivariant under this action, i.e.,
$\gam^\cO(\{f,g\})=\{\gam^\cO(f),\gam^\cO(g)\}$ for all $\gam\in\SO(V,q)$
and all fermionic functions $f,g$ on $\PV$.

\section{A family of star products of fermionic functions}\label{sec:starK}
Consider a linear fermionic system of finite degrees of freedom, whose phase
space $\PV$ is a fermionic copy of the Euclidean space $(V,q)$.
Two fermionic functions functions $f,g\in\extb V^*_\bC$ have a Moyal-type
star product \cite{Bo,Za,HH}
\begin{equation}\label{eqn:moyal}
f*_0g=f\,\big(e^{-\frac\hbar4q^\mn\lpdr_\mu\rpdr_\nu}\big)\,g
     =f\wedge g+\mfrac\hbar2\,\{f,g\}+O(\hbar^2).
\end{equation}
Here the Plank constant $\hbar>0$ is fixed.
Since \eqref{eqn:moyal} is a finite sum, the fermionic star product is always
non-formal.
Using the notation in \S\ref{sec:poisson}, the star product can also be
written as
\[ f*_0g=\Del^*(e^{-\frac\hbar4q^\mn\pdm\hot\pdn}f\hot g\big).  \]
This star product is associative but not graded commutative, and it satisfies,
for all fermionic functions $f,g$ on $\PV$,
\[  f*_0g-(-1)^{|f||g|}g*_0f=\hbar\,\{f,g\}+O(\hbar^2).  \]
For example, $1*_0f=f*_01=f$ for all $f$ and
$a*_0b=a\wedge b+\frac\hbar4q^\sharp(a,b)$ if $a,b\in V^*_\bC$ are regarded as
fermionic functions on $\PV$.
It is easy to show that for any $a=a_\mu e^\mu\in V^*_\bC$ and
$f,g\in\extb V^*_\bC$, we have
\begin{equation}\label{eqn:awedgef}
\begin{aligned}
(a\wedge f)*_0g
&=a\wedge(f*_0g)+(-1)^{|f|}\qt\hbar\,q^\mn a_\mu f*_0(\pdn g),        \\
f*_0(a\wedge g)
&=(-1)^{|f|}a\wedge(f*_0g)+\qt\hbar\,q^\mn a_\mu(\pdn f)*_0g.
\end{aligned}
\end{equation}

We now introduce a family of fermionic star products parametrised by
anti-symmetric contravariant $2$-tensors $K$;
in the bosonic case, $K$ would be symmetric \cite{OMMY05,OMMY07}.
For each such $K\in\ext2V_\bC$, let $\Lam:=q^\sharp+K\in V_\bC^{\otimes2}$ or
$\Lam^\mn=q^\mn+K^\mn$ in components, and let the corresponding (non-formal)
star product $*_K$ be defined by
\begin{equation}\label{eqn:stark}
f*_Kg:=f\,\big(e^{-\frac\hbar4\Lam^\mn\lpdr_\mu\rpdr_\nu}\big)\,g
     =\Del^*(e^{-\frac\hbar4\Lam^\mn\pdm\hot\pdn}f\hot g\big).
\end{equation}
Then we have $1*_Kf=f*_K1=f$ and $f*_Kg=f\wedge g+O(\hbar)$ for all $f,g$ as
before, but for $a,b\in V^*_\bC$,
\[  a*_Kb=a\wedge b+\mfrac\hbar4\Lam(a,b)=a*_0b+\mfrac\hbar4K(a,b).  \]
When $K=0$, $*_K=*_0$ is the fermionic Moyal star product \eqref{eqn:moyal}.
Since $K$ is anti-symmetric, we still have
\begin{equation}\label{eqn:first}
f*_Kg-(-1)^{|f||g|}g*_Kf
=-2\Del^*\big(\sinh\big(\tfrac\hbar4q^\mn\pdm\hot\pdn\big)\,
 e^{\frac\hbar4K^\mn\pdm\hot\pdn}(f\hot g)\big)=\hbar\,\{f,g\}+O(\hbar^2).
\end{equation}
Like \eqref{eqn:moyal}, the star products \eqref{eqn:stark} are polynomials
in $\hbar$ of degree at most $m=\dim V$.
So there is no problem of convergence that should be taken care of in the
bosonic case \cite{WY}.

We verify the associativity of the fermionic star product $*_K$ for a general
$K\in\ext2V_\bC$.
Using the Leibniz rule \eqref{eqn:leib} and property \eqref{eqn:diag} of
the diagonal map, we obtain, for all fermionic functions $f,g,h$ on $\PV$,
\begin{align*}
(f*_Kg)*_Kh&=\Del^*\big(e^{-\frac\hbar4\Lam^\mn\pdm\hot\pdn}\big(\Del^*
 \big(e^{-\frac\hbar4\Lam^\mn\pdm\hot\pdn}f\hot g\big)\hot h\big)\big)   \\
&=\Del^*(\Del\times1)^*\big(e^{-\frac\hbar4\Lam^\mn(\pdm\hot1+1\hot\pdm)
 \hot\pdn-\frac\hbar4\Lam^\mn\pdm\hot\pdn\hot1}f\hot g\hot h\big)        \\
&=\Del_{(3)}^*\big(e^{-\frac\hbar4\Lam^\mn
 (1\hot\pdm\hot\pdn+\pdm\hot1\hot\pdn+\pdm\hot\pdn\hot1)}f\hot g\hot h\big).
\end{align*}
A similar calculation shows that $f*_K(g*_Kh)$ is the same, and thus
associativity
\begin{equation}\label{eqn:assoc}
(f*_Kg)*_Kh=f*_K(g*_Kh)
\end{equation}
follows.
By the first order behaviour \eqref{eqn:first} of the star product $*_K$,
the graded Jacobi identity \eqref{eqn:jacobi} is a consequence of associativity
\eqref{eqn:assoc}.

We now consider the star products $*_K$ parametrised by $K\in\ext2V_\bC$ as a
family.
Let $\cO$ be the product bundle over $\ext2V_\bC$ whose fibre over $K$ is the
same space $\extb V^*_\bC$ of fermionic functions but with star product $*_K$.
Following a similar construction in bosonic theory \cite{OMMY05,OMMY07} (see
\cite{AW,Ba,Za,OMMY03} for special cases of normal orderings), for any
$K,K'\in\ext2V_\bC$, we define an intertwining operator $\U^\cO_{K',K}$ on
$\extb V^*_\bC$ (identified as the fibres over $K$ and $K'$) by
\[ \U^\cO_{K',K}f:=e^{-\frac\hbar8(K'-K)^\mn\pdm\pdn}f.    \]
Clearly, $\U^\cO_{K'',K}=\U^\cO_{K'',K'}\circ\U^\cO_{K',K}$ for any three
$K,K',K''\in\ext2V_\bC$.
We verify that $\U^\cO_{K',K}$ intertwines the star products $*_K$ and
$*_{K'}$.
Indeed, for all fermionic functions $f,g$ on $\PV$,
\begin{align*}
\U^\cO_{K',K}(f*_Kg)&=e^{-\frac\hbar8(K'-K)^\mn\pdm\hot\pdn}\;
  \Del^*\big(e^{-\frac\hbar4\Lam^\mn\pdm\pdn}f\hot g\big)                   \\
&=\Del^*\big(e^{-\frac\hbar8(K'-K)^\mn(\pdm\hot1+1\hot\pdm)
  (\pdn\hot1+1\hot\pdn)-\frac\hbar4(q^\sharp+K)^\mn\pdm\hot\pdn}f\hot g\big)\\
&=\Del^*\big(e^{-\frac\hbar4(q^\sharp+K')\pdm\hot\pdn}(e^{-\frac\hbar8
  (K'-K)^\mn\pdm\pdn}f)\hot(e^{-\frac\hbar8(K'-K)^\mn\pdm\pdn}g)\big)       \\
&=(\U^\cO_{K',K}f)*_{K'}(\U^\cO_{K',K}g).
\end{align*}

As $K\in\ext2V_\bC$ varies by $\del K$, the change in $f\in\extb V^*_\bC$
under the intertwining operator is, to the first order,
\[ \del f=(\U^\cO_{K+\del K,K}-1)f=-\mfrac\hbar8\,\del K^\mn\pdm\pdn f.   \]
This defines a connection $\nabla^\cO$ in the product bundle $\cO$: as a
point $K$ in $\ext2V_\bC$ moves to $K+\del K$, the infinitesimal parallel
transport is $f\mapsto f+\del f$.
Thus along a smooth path $\{K_t\}_{0\le t\le1}$ in $\ext2V_\bC$, the parallel
transport from $K_0$ to $K_1$ is $f\mapsto\U^\cO_{K_1,K_0}f$.
The connection $\nabla^\cO$ is flat because $\U^\cO_{K_1,K_0}f$ depends only
on the end points $K_0,K_1$ of the path.
Regarding each $f\in\extb V^*_\bC$ as a constant section of $\cO$, the
connection $1$-form on $\ext2V_\bC$ is
\[ A^\cO=\mfrac\hbar8\,(\del K)^\mn\pdm\pdn.  \]
A direct calculation confirms that the curvature $F^\cO=0$.
The connection $\nabla^\cO$ respects the star products as the parallel
transport $\U^\cO_{K_1,K_0}$ intertwines between $*_{K_0}$ and $*_{K_1}$.
Infinitesimally, this is reflected in the first order identity
\[ f\,(*_{K+\del K}-*_K)\,g=\del(f*_Kg)-(\del f)*_Kg-f*_K(\del g).  \]

Both the star product $*_K$ and the intertwining operator $\U^\cO_{K',K}$
are equivariant under the action of $\SO(V,q)$.
That is, for all $\gam\in\SO(V,q)$, $K,K'\in\ext2V_\bC$ and fermionic
functions $f,g$ on $\PV$, we have
\[ \gam^\cO(f*_Kg)=\gam^\cO(f)*_{(\gam\otimes\gam)K}\gam^\cO(g),\qquad
\gam^\cO(\U^\cO_{K',K}f)=\U^\cO_{(\gam\otimes\gam)K',(\gam\otimes\gam)K}
(\gam^\cO(f)).                                                         \]
In particular, the fermionic Moyal product $*_0$ is invariant under
$\SO(V,q)$.
The action of $\SO(V,q)$ on $\ext2V_\bC$ lifts to the bundle $\cO$ via
$\gam\colon(K,f)\mapsto((\gam\otimes\gam)K,\gam^\cO(f))$.
Since $\U^\cO_{K',K}$ is the parallel transport along any smooth path from $K$
to $K'$, the lifted action of $\SO(V,q)$ on $\cO$ preserves the flat
connection $\nabla^\cO$.

\section{Polarisations, polarised star products and normal ordering}
\label{sec:polar}
Let $(V,q)$ be a Euclidean space of even dimension $m$.
A complex subspace $W\subset V_\bC$ is isotropic if $W\subset W^\perp$
and is coisotropic if $W^\perp\subset W$.
For simplicity, we will consider mostly the case when the dimension $m=2n$
is even.
Then the isotropic subspaces of maximal dimension coincides with the
coisotropic subspaces of minimal dimension; these are the complex
Lagrangian subspaces $L\subset V_\bC$ satisfying $L=L^\perp$.

If $(L,L')$ is a pair of transverse complex Lagrangian subspaces, we have
the linear decomposition $V_\bC=L\oplus L'$.
The projection operator $P$ on $V_\bC$ such that $\im P=L$ and $\ker P=L'$
satisfies the properties
\begin{equation}\label{eqn:projection}
P^2=P,\qquad(P\otimes P)q=0,\qquad((1-P)\otimes(1-P))q=0.
\end{equation}
Conversely, given an operator $P$ on $V_\bC$ satisfying the above properties
\eqref{eqn:projection}, we can recover the pair $(L,L')$ of transverse
complex Lagrangian subspaces in $V_\bC$.
Therefore the space $\cP$ of operators $P$ satisfying \eqref{eqn:projection}
naturally identifies with the space of pairs $(L,L')$ and is a complex
manifold of dimension $\hf n(n-1)$.
Choosing linear bases $\{e_i\}_{1\le i\le n}$ of $L$ and
$\{e_{i'}\}_{1\le i'\le n}$ of $L'$, the non-zero components of $q$ and
$q^\sharp$ are, respectively, $q_{i'j}=q_{ji'}$ and $q^{i'j}=q^{ji'}$.

Given a projection operator $P\in\cP$ or a pair of $(L,L')$ transverse
Lagrangian subspaces, we set
\[ K_P:=((1-P)\otimes P-P\otimes(1-P))q\in\ext2V_\bC,\qquad
   \Lam_P:=q^\sharp+K_P=2((1-P)\otimes P)q^\sharp. \]
The map $\cP\to\ext2V_\bC$, $P\mapsto K_P$ is injective and thus $\cP$ can be
regarded as an analytic subset of $\ext2V_\bC$.
With the linear bases of $L$ and $L'$, the non-zero components of $K_P$
and $\Lam_P$ are, respectively, $K_P^{i'j}=-K_P^{ji'}=q^{i'j}$ and
$\Lam_P^{i'j}=2q^{i'j}$.

By \eqref{eqn:projection}, an infinitesimal change $\del P$ in $P\in\cP$
satisfies the constraints
\[ P\,\del P=\del P\,(1-P),\quad(1-P)\,\del P=\del P\,P,\quad
   (\del P\otimes P+P\otimes\del P)q^\sharp=0,\quad
   (\del P\otimes(1-P)+(1-P)\otimes\del P)q^\sharp=0.   \]
It follows that $(\del P\otimes1+1\otimes\del P)q^\sharp=0$ and thus
$(\del P\otimes1)q^\sharp\in\ext{2}V_\bC$ is anti-symmetric.
Under the above linear bases, the non-zero components of $\del P$ are
$(\del P)^i_{\;j'}$ and $(\del P)^{i'}_{\;j}$, and those of
$(\del P\otimes1)q^\sharp$ are
$(\del P)^{ij}=(\del P)^i_{\;k'}q^{k'j}=-(\del P)^{ji}$,
$(\del P)^{i'j'}=(\del P)^{i'}_{\;k}q^{kj'}=-(\del P)^{j'i'}$.
If $P$ varies by $\del P$, the change in $K_P\in\ext2V_\bC$ is
$\del K_P=(-\del P\otimes1+1\otimes\del P)q^\sharp=-2(\del P\otimes1)q^\sharp$.
Thus we obtain
\[ (P\otimes P)\del K_P=-2(\del P\otimes P)q^\sharp,\qquad
   ((1-P)\otimes(1-P))\del K_P=-2(\del P\otimes(1-P))q^\sharp,   \]
or $(\del K_P)^{ij}=-2(\del P)^{ij}$, $(\del K_P)^{i'j'}=-2(\del P)^{i'j'}$
in components.
Clearly, $\del\Lam_P=\del K_P$.

An important part of $\cP$ is the set $\cJ$ of linear complex structures
$J$ on $V$ that is compatible with the orientation on $V$ and such that
$q(J\cdot,J\cdot)=q(\cdot,\cdot)$ or $(J\otimes J)q^\sharp=q^\sharp$.
For each $J\in\cJ$, the $(1,0)$- and $(0,1)$-subspaces $L_J:=V_J^{1,0}$ and
$\overline{L_J}=V_J^{0,1}$ form a pair of transverse Langrangian subspaces
in $V_\bC$.
The projection onto $L_J$ is $P_J:=\hf(1-\ii J)\in\cP$, and
$\overline{P_J}=1-P_J$ is the projection onto $\overline{L_J}$.
This defines an inclusion map $\si\colon\cJ\hookrightarrow\cP$, $J\mapsto P_J$.
The space $\cJ$ is a compact Hermitian symmetric space of complex dimension
$\hf n(n-1)$ of type $\SO(2n)/\rU(n)$, and is connected and simply connected.
If $n=1$, $\cJ$ is a point.
If $n>1$, then $H_2(\cJ)\cong\pi_2(\cJ)\cong\bZ$.
The standard K\"ahler form on $\cJ$ is
\begin{equation}\label{eqn:upom}
\upom_q=\ii\tr(P_J\del P_J\wedge\del P_JP_J)
   =-\tfrac\ii4\tr(P_J\del J\wedge\del JP_J),
\end{equation}
whose integration on the generator of $H_2(\cJ)$ or $\pi_2(\cJ)$ is $4\pi$.
For example, $\cJ=\bP^1$ if $n=2$, and the K\"ahler form is
\[ \upom_q=\mfrac{2\ii dz\wedge d\bar z}{(1+|z|^2)^2}  \]
in the inhomogeneous coordinate $z\in\bC\cup\{\infty\}$.
In general, for two complex structures $J,J'\in\cJ$, the condition that
$\overline{L_J}$ and $L_{J'}$ are transverse to each other is $\det(J+J')\ne0$.
This happens if and only if $J'$ is not on the cut locus or the first
conjugate locus of $J$ in $\cJ$ \cite{Wu15}.
For a fixed $J\in\cJ$, the set of such $J'$ is diffeomorphic to
$\bC^{n(n-1)/2}$.
When $n=2$, $J'$ is on the cut locus of $J$ if $J$ and $J'$ are antipodal
points on $\bP^1=S^2$.

There is a retraction map $\sr\colon\cP\to\cJ$, sending $P\in\cP$ to
the $J\in\cJ$ such that $P_J=P(P-\oP)^{-1}(1-\oP)$, or
\[ \sr\colon P\mapsto J=\ii(P(P-\oP)^{-1}(1-\oP)+\oP(P-\oP)^{-1}(1-P)).  \]
Here $P-\oP$ is invertible because $q$ is positive definite on the real space
$V$ and hence $\im P$ and $\im\oP$ are necessarily transverse subspaces.
It is easy to verify that $\sr\circ\si=\id_\cJ$.
In fact, $J=\sr(P)$ picks up only the information of $\im P$ but not of
$\ker P$.
With $J'=\sr(1-P)$, $\cP$ can be identified with an open dense subset of
$\cJ\times\cJ$ consisting of pairs $(J,J')$ such that $J'$ is not on the
cut locus of $-J$.
Since $J'$ can be deformed to $-J$ following the unique minimal geodesic
joining $J'$ to $-J$, the subspace $\cJ$ is a strong deformation retract
of $\cP$.
Moreover, $\sr\colon\cP\to\cJ$ is a smooth (but not holomorphic) fibration
whose fibres are diffeomorphic to $\bC^{n(n-1)/2}$.
The other map $\sr'\colon\cP\to\cJ$, $P\mapsto J'$, has similar properties.
Consequently, the homology and homotopy groups of $\cP$ coincide with those
of $\cJ$.
In particular, $\cP$ is connected and simply connected, and
$H_2(\cP)\cong\pi_2(\cP)\cong\bZ$ just like $\cJ$.

If $P\in\cP$, the polarised star product $*_P$ (with respect to $P$) is the
star product $*_K$ in \eqref{eqn:stark} for $K=K_P$.
That is, for two fermionic functions $f,g\in\extb V^*_\bC$, we have
\begin{equation}\label{eqn:starP}
f*_Pg=f\,\big(e^{-\frac\hbar2\,q^{i'j}\lpdr_{i'}\rpdr_j}\big)\,g
=f\wedge g+\mfrac\hbar2\,q^{i'j}\pdr_{i'}f\wedge\pdr_j g+O(\hbar^2).
\end{equation}
In particular, if $\pdr_{i'}f=0$ or $\pdr_jg=0$, then $f*_Pg=f\wedge g$.
This star product is related to the normal ordering with respect to $P$.
(See for example \cite{AW,Ba,Za,OMMY03} for the bosonic analogue.)
If $P,P'\in\cP$, we write $\U^\cO_{P',P}$ for $\U^\cO_{K_{P'},K_P}$.
If $J,J'\in\cJ$, we write $*_J$ for $*_{P_J}$ 
and $\U^\cO_{J'J}$ for $\U^\cO_{P_{J'},P_J}$.

The group $\SO(V,q)$ acts on pairs of transverse Lagrangian subspaces
$(L,L')$ by $\gam\colon(L,L')\mapsto(\gam L,\gam L')$, where
$\gam\in\SO(V,q)$.
The action on the corresponding projection operator $P\in\cP$ is
$\gam\colon P\mapsto\gam\circ P\circ\gam^{-1}$.
Similarly, $\SO(V,q)$ acts on a complex structure $J\in\cJ$ by
$\gam\colon J\mapsto\gam\circ J\circ\gam^{-1}$.
The inclusion maps $\cP\to\ext2V_\bC$ and $\si\colon\cJ\to\cP$ are
equivariant under the actions of $\SO(V,q)$, and so are the retraction maps
$\sr,\sr'\colon\cP\to\cJ$.
The polarised star product 
is also equivariant under $\SO(V,q)$, i.e.,
$\gam^\cO(f*_Pg)=\gam^\cO(f)*_{\gam P\gam^{-1}}\gam^\cO(g)$
for all $\gam\in\SO(V,q)$ and fermionic functions $f$ and $g$ on $\PV$.
We also have $\gam^\cO(\U^\cO_{P',P}(f))=
\U^\cO_{\gam P'\gam^{-1},\gam P\gam^{-1}}(\gam^\cO(f))$ for $P,P'\in\cP$.
In particular, taking $P=P_J$, $P'=P_{J'}$ with $J,J'\in\cJ$, then
$\gam^\cO(f*_Jg)=\gam^\cO(f)*_{\gam J\gam^{-1}}\gam^\cO(g)$
and $\gam^\cO
(\U^\cO_{J',J}(f))=\U^\cO_{\gam J'\gam^{-1},\gam J\gam^{-1}}(\gam^\cO(f))$.

\section{Prequantisation, quantisation, projective flatness and symmetry}
\label{sec:qtn}
Quantisation of the fermionic phase space $(V,q)$ means finding an
irreducible representation of the operator algebra, the Clifford algebra $\Cl$.
It is well known that such a representation is the spinor representations that
can be constructed with the help of a complex structure on $V$ \cite{Ch}.
In the procedures of prequantisation \cite{Ko77} and quantisation \cite{Wo}
of a fermionic system which lead to the spinor representation as the quantum
Hilbert space, the complex structure plays the role of polarisation.
Just like the bosonic case \cite{ADPW,KW}, there is a bundle of spinor
representations over the space $\cJ$ of complex structures and this bundle
admits a projectively flat connection allowing natural identification up to
a phase of the spinor representation from different complex structures
\cite{Wu15,Wu17}.
Now we extend the base space $\cJ$ of the projectively flat bundle to the
space $\cP$ of projections $P$ satisfying \eqref{eqn:projection}.

The prequantum line bundle $\ell$ over $\PV$ does not have a geometric total
space but is characterised by the space $\Gam(\PV,\ell)$ of sections and the
operators acting on them.
The space $\Gam(\PV,\ell)$ is a free left module of rank $1$ over
$\extb V^*_\bC$, with a multiplication $f\psi\in\Gam(\PV,\ell)$ such that
$(f\wedge g)\psi=f(g\psi)$ for all $f,g\in\extb V^*_\bC$ and
$\psi\in\Gam(\PV,\ell)$.
Furthermore it has an Hermitian structure and a unitary connection.
The covariant derivative $\nabla_v$ along a real vector $v\in V$ acts on
$\Gam(\PV,\ell)$ as a skew-Hermitian operator satisfying
$\nabla_v(f\psi)=(-1)^{|f|}f\nabla_v\psi+(\pdr_vf)\psi$ for all
$f\in\extb V^*_\bC$ and $\psi\in\Gam(\PV,\ell)$.
It is further required to have the (symmetric) curvature $2$-form
\begin{equation}\label{eqn:curvl}
\{\nabla_u,\nabla_v\}=-2\hbar^{-1}q(u,v),
\end{equation}
where $u,v\in V_\bC$.
The prequantum action of a fermionic function $f\in\extb V^*_\bC$ on
$\Gam(\PV,\ell)$ is
\begin{equation}\label{eqn:preqtm}
\hat f=f+\hf\hbar\,\nabla_{H_f},
\end{equation}
where $H_f$ is the Hamiltonian vector field \eqref{eqn:ham} of $f$.
The prequantum action satisfies the Dirac condition
\begin{equation}\label{eqn:dirac}
[\hat f,\hat g]=\hbar\,\widehat{\{f,g\}}
\end{equation}
for all $f,g\in\extb V^*_\bC$, where $[\hat f,\hat g]$ is the graded
commutator of the prequantum operators $\hat f$ and $\hat g$.

We assume that the dimension of $V$ is $m=2n$.
Given $P\in\cP$ or a pair $(L,L')$ of transverse complex Lagrangian subspaces,
we set $\cH_P:=\Gam_P(\PV,\ell)$ as the space of sections
$\psi\in\Gam(\PV,\ell)$ that are covariantly constant along $L'=\ker P$, i.e.,
$\nabla_v\psi=0$ for all $v\in L'$.
This is the quantum Hilbert space in the polarisation given by $P$ and is of
complex dimension $2^n$.
For $a\in V^*_\bC$ regarded as an element of $\extb V^*_\bC$, the prequantum
action $\hat a$ preserves $\cH_P$ and for $a,b\in V^*_\bC$, we have the
relation $\{\hat a,\hat b\}=\frac\hbar2q^\sharp(a,b)$.
This induces the (irreducible) spinor representation of the Clifford algebra
$\Cl$ on $\cH_P$.
Furthermore, there is a decomposition
\begin{equation}\label{eqn:decomp}
\Gam(\PV,\ell)=\cH_P\oplus\cH'_P,
\end{equation}
where $\cH'_P$ is the subspace spanned by sections of the form $\nabla_v\psi'$
for $v\in L$ and $\psi'\in\Gam(\PV,\ell)$.
Note that $\cH_P$ depends only on $L'$ whereas $\cH'_P$ depends only on $L$,
but the decomposition \eqref{eqn:decomp} depends on the full information of
$P\in\cP$.
If $P=P_J$ is from a complex structure $J\in\cJ$ or if $L=V_J^{1,0}$,
$L'=V_J^{0,1}$, then $\nabla_{\bar v}$ is the adjoint of $-\nabla_v$ for all
$v\in L$ and thus the decomposition \eqref{eqn:decomp} is orthogonal with
respect to the Hermitian structure on $\Gam(\PV,\ell)$ \cite{Wu15}.

Choosing a linear basis $\{e_\mu\}$ of $V$, we have fermionic coordinates
$\{\tht^\mu\}$ on $V$.
A `trivialisation' of $\ell$ identifies the $\extb V^*_\bC$-module
$\Gam(\PV,\ell)$ with $\extb V^*_\bC$, under which the covariant derivative
$\nabla_\mu$ becomes the usual derivative plus a gauge potential with the
curvature condition \eqref{eqn:curvl}.
For example, the choice $\nabla_\mu=\pdm-\hbar^{-1}q_\mn\tht^\nu$ satisfies
$\{\nabla_\mu,\nabla_\nu\}=-2\hbar^{-1}q_\mn$.
Given linear bases $\{e_i\}$ of $L=\im P$ and $\{e_{i'}\}$ of $L'=\ker P$, we
have complex coordinates $\tht^i$ and $\tht^{i'}$.
The space $\cH_P$ consists of $\psi\in\Gam(\PV,\ell)$ satisfying
$\nabla_{i'}\psi=0$; such $\psi$ is of the form
$\phi(\tht^i)\exp(\hbar^{-1}q_{i'j}\tht^{i'}\tht^j)$, where
$\phi\in\extb L^*$ depends only on $\tht^i$ but not on $\tht^{i'}$.
Notice here the appearance of the fermionic Gaussian factor.
On the other hand the space $\cH'_P$ consists of sections of the form
$\nabla_i\psi^i$ for some $\psi^i\in\Gam(\PV,\ell)$.
If $P=P_J$ for $J\in\cJ$, then $L'=\overline L$ and we can choose $e_{i'}$ as
the complex conjugate of $e_i$ and write $\bar i$ for $i'$.
Then $\tht^i$ are the holomorphic coordinates whereas $\tht^{\bar i}$ are the
anti-holomorphic coordinates, and $\phi(\tht^i)$ is a (fermionic) holomorphic
function.

The finite dimensional Hilbert spaces $\cH_P$ parametrised by $P\in\cP$ form
a vector bundle $\cH$ over $\cP$ and it is a subbundle of the product bundle
with fibre $\Gam(\PV,\ell)$.
Similarly, $\cH'_P$ form a subbundle $\cH'$ which is complement to $\cH$
in the trivial bundle.
The trivial connection on the product bundle then induces a connection
$\nabla^\cH$ on $\cH$.
Under an infinitesimal variation $\del P$ of $P\in\cP$, the change in
$\psi\in\cH_P$ by parallel transport is
\[ \del\psi=-\mfrac14\nabla_i(\del P)^{ij}\nabla_j\psi
   =-\mfrac14(\del P)^{ij}\nabla_i\nabla_j\psi.    \]
Clearly, $\del\psi$ is valued in $\cH'_P$ and, following
\cite[\S III.B]{Wu15}, it can be verified that to the first order,
$\psi+\del\psi$ is in $\cH_{P+\del P}$.
The connection $1$-form of $\nabla^\cH$ is
$\A^\cH=\qt(\del P)^{ij}\nabla_i\nabla_j$ and its curvature is
(by the same calculations in \cite[\S III.B]{Wu15})
\begin{equation}\label{eqn:curvH}
\F^\cH=-\hf\tr(P\del P\wedge\del PP)\id_\cH.
\end{equation}

Since $\F^\cH$ is a $2$-form on $\cP$ times the identity section $\id_\cH$ of
$\End(\cH)$, the connection $\nabla^\cH$ is projectively flat.
Along a smooth path $\{P_t\}_{0\le t\le1}$ in $\cP$, let
$\U^\cH_{\{P_t\}}\colon\cH_{P_0}\to\cH_{P_1}$ be the parallel transport
in $\cH$.
If $\{P'_t\}$ is another path connecting the same points $P_0$ and $P_1$,
then $\U^\cH_{\{P_t\}}$ and $\U^\cH_{\{P'_t\}}$ differ by a phase.
So unlike $\U^\cO_{P',P}$, the notation $\U^\cH_{P',P}$ has a phase ambiguity.
The spaces $\cH_P$ of spinors constructed from various $P\in\cP$ can be
identified naturally up to a phase.
Restricting to the subset $\cJ$ of the base space $\cP$, the decomposition
\eqref{eqn:decomp} is orthogonal and hence the connection $\nabla^\cH$ is
given by the orthogonal projection of the trivial connection onto the
subbundle $\cH$ \cite{Wu15}.
The connection $\nabla^\cH$ over $\cJ$ is unitary and the curvature
\eqref{eqn:curvH} is proportional to the standard K\"ahler form on the
Hermitian symmetric space $\cJ$.
If $\{J_t\}_{0\le t\le1}$ is a smooth path in $\cJ$, we write
$\U^\cH_{\{J_t\}}$ for $\U^\cH_{\{P_{J_t}\}}$, which is unitary.

The action of the symmetry group $\SO(V,q)$ on the phase space $\PV$ can be
lifted to the prequantum line bundle $\ell$ in the sense that it acts on the
space $\Gam(\PV,\ell)$ of sections.
Since action preserves the Hermitian structure and the unitary connection
$\nabla$ on $\ell$, it maps $\cH_P$ to $\cH_{\gam P\gam^{-1}}$ and thus the
action of $\SO(V,q)$ on $\cP$ lifts to the bundle $\cH$.
Let $\gam^\cH$ denote the action of $\gam\in\SO(V,q)$ on $\cH$.
The lifted action preserves the projectively flat connection $\nabla^\cH$.
If $\{P_t\}_{0\le t\le1}$ is a smooth path in $\cP$, the parallel transport
in $\cH$ satisfies
\[ \gam^\cH\circ\U^\cH_{\{P_t\}}
   =\U^\cH_{\{\gam P_t\gam^{-1}\}}\circ\gam^\cH.\]

The group $\SO(V,q)$ acts as automorphisms of the Clifford algebra $\Cl$.
When a classical symmetry becomes part of the automorphism group of the
quantum operator algebra, it is said to be broken in an irreducible
representation of the operator algebra if the symmetry does not preserve
the representation.
In our case, if $\dim V$ is even, since the irreducible representation of
$\Cl$ is unique (the fermionic analogue of the Stone-von Neumann theorem),
the $\SO(V,q)$ symmetry is necessarily unbroken.
In fact, picking any $P\in\cP$, a projective representation $\rho_P$ of
$\SO(V,q)$ on $\cH_P$ can be constructed by
$\rho_P(\gam)=\U^\cH_{P,\gam P\gam^{-1}}\circ\gam^\cH$, which is defined up
to a phase.
Indeed, we have \cite{Wu17}
\begin{align*}
\rho_P(\gam\gam')
&=\U^\cH_{P,(\gam\gam')P(\gam\gam')^{-1}}\circ(\gam\gam')^\cH
=\U^\cH_{P,\gam P\gam^{-1}}\circ
\U^\cH_{\gam P\gam^{-1},\gam(\gam'P\gam'{}^{-1})\gam^{-1}}\circ\gam^\cH
\circ(\gam')^\cH                                                        \\
&\qquad\qquad
=\U^\cH_{P,\gam P\gam^{-1}}\circ\gam^\cH\circ\U^\cH_{P,\gam'P\gam'{}^{-1}}
\circ(\gam')^\cH=\rho_P(\gam)\circ\rho_P(\gam')
\end{align*}
up to a phase for all $\gam,\gam'\in\SO(V,q)$.
So $\rho_P$ is a projective representation of $\SO(V,q)$ which lifts to an
honest representation of its double cover, the spin group $\Spin(V,q)$, and
it is unitary if we pick $P=P_J$, $J\in\cJ$.
This is the fermionic analogue of the Segal-Shale-Weil representation of the
metaplectic group (the double cover of the symplectic group).
More generally, the symmetry group acts on the space of polarisations.
The symmetry is unbroken if one can find a polarisation such that over
its orbit, the bundle of Hilbert spaces is projectively flat \cite{Wu17}.

Actually the automorphism group of $\Cl$ contains the full orthogonal group
$\Or(V,q)$.
When $\dim V$ is even as we have been considering, the full $\Or(V,q)$
symmetry is unbroken because parity, in the non-identity component of
$\Or(V,q)$, remains an inner automorphism of $\Cl$ and it acts on the spinors
by exchanging positive and negative chiralities.
However, when $\dim V$ is odd, though the $\SO(V,q)$ part remains inner,
parity becomes an outer automorphism of $\Cl$ and it exchanges the two
irreducible representations of $\Cl$.
Therefore the $\Or(V,q)$ symmetry is broken to $\SO(V,q)$ in each
irreducible spinor representation when $\dim V$ is odd \cite{Wu17}.

\section{Star product of a fermionic function on a state}\label{sec:starP}
Given a polarisation $P\in\cP$ on the fermionic phase space $(V,q)$, the
star products \eqref{eqn:starP} of two fermionic functions
$f,g\in\extb V_\bC^*$ is normal ordered with respect to $P$.
On the other hand, $P\in\cP$ also determines the space $\cH_P=\Gam_P(V,\ell)$
of polarised sections $\psi$ of the prequantum line bundle $\ell$ satisfying
$\nabla_{i'}\psi=0$.
As in the bosonic case \cite{WY,Wu19}, it is desirable to have a representation
of the algebra with the star product $*_P$ on the space $\cH_P$ of quantum
states defined by the polarisation $P$.
Indeed, we define the star product $*_P$ of a function $f\in\extb V_\bC^*$ on
a section $\psi\in\Gam(\PV,\ell)$ by
\begin{equation}\label{eqn:f*psi}
f*_P\psi=\Del^*\big(e^{-\frac14\hbar\,\Lam_P(d\otimes\nabla)}f\otimes\psi\big)
=\Del^*\big(e^{-\frac12\hbar\,q^{i'j}\pdr_{i'}\otimes\nabla_j}
 f\otimes\psi\big)
=f\,e^{-\frac12\hbar\,q^{i'j}\lpdr_{i'}\rnabla_j}\,\psi.
\end{equation}
Here we regard $f\otimes\psi\in\extb V_\bC^*\otimes_\bC\Gam(\PV,\ell)$ as a
section of the line bundle $1\boxtimes\ell=\pi_2^*\ell$ over $\PV\times\PV$
(where $\pi_2\colon V\times V\to V$ is the projection onto the second factor),
whose restriction to the diagonal is $\Del^*(1\boxtimes\ell)\cong\ell$.
If $\psi\in\cH_P$ is polarised (or in the quantum Hilbert space defined by
the polarisation $P$), we have
\begin{equation}\label{eqn:f*psi0}
f*_P\psi=f\psi-\hf\hbar\,q^{i'j}\pdr_{i'}f\,\nabla_j\psi+o(\hbar^2)
        =f\psi+\hf\hbar\,\nabla_{H_f}\psi+o(\hbar^2),
\end{equation}
where $H_f=q^{ij'}(\pdr'_{j'}f)\pdr_i+q^{i'j}(\pdr'_jf)\pdr_{i'}$ is the
Hamiltonian vector field \eqref{eqn:ham} of $f$ on $\PV$.
The first two terms in \eqref{eqn:f*psi0} are precisely the prequantum action
\eqref{eqn:preqtm} of $f$ on $\psi$.
Thus the star product defined here is a deformation of prequantisation.
If $f$ is constant or linear in the $\tht^{i'}$ variables, then the remaining
terms in \eqref{eqn:f*psi0} vanish and the $*_P$-product coincides with the prequantum action of $f$ on $\cH_P$.

The star product $*_P$ is associative in the sense that
$f*_P(g*_P\psi)=(f*_Pg)*_P\psi$ for $f,g\in\extb V_\bC^*$ and
$\psi\in\Gam(\PV,\ell)$.
To verify this equality, we formulate the Leibniz rule
$\nabla(f\psi)=f\nabla\psi+df\otimes\psi$ as an operator identity
$\nabla\circ\Del^*=\Del^*\circ(d\otimes1+1\otimes\nabla)$ like
\eqref{eqn:leib} but acting on $\extb V_\bC^*\otimes_\bC\Gam(\PV,\ell)$.
Note that
$(((1-P)\otimes P)q^\sharp)(d\otimes\nabla)=q^{i'j}\pdr_{i'}\otimes\nabla_j$
contains $\nabla_j$ but not $\nabla_{i'}$.
Since $\{\nabla_i,\nabla_j\}=0$, the operators that appear as exponents in
the star product $*_P$ anti-commute with each other just as in the case for
functions.
Thus similar calculations show that both $f*_P(g*_P\psi)$ and $(f*_Pg)*_P\psi$
are
\[ \Del_{(3)}^*\big(e^{-\frac12\hbar\,q^{i'j}
   (1\otimes\pdr_{i'}\otimes\nabla_j+\pdr_{i'}\otimes1\otimes\nabla_j
   +\pdr_{i'}\otimes\nabla_j\otimes1)}f\otimes g\otimes\psi\big).     \]
As a consequence of associativity, we have
\[ f*_P(g*_P\psi)-g*_P(f*_P\psi)=\hbar\,\{f,g\}*_P\psi+o(\hbar^2),  \]
which is consistent with the appearance of the prequantum action in $f*_P\psi$.

Next, we show that if $\psi\in\cH_P$, then $f*_P\psi\in\cH_P$ for all
$f\in\extb V_\bC^*$.
This is somewhat surprising, as the function $f$ is not required to be
holomorphic (in the fermionic sense).
Yet it is a crucial requirement for the star product \eqref{eqn:f*psi}:
the action of an observable on a quantum state must remain a quantum state.
Indeed, since $\nabla_{i'}\psi=0$, we obtain
\begin{align*}
\nabla_{i'}(f*_P\psi)&=\Del^*\big((\pdr_{i'}\otimes1+1\otimes\nabla_{i'})
  \,e^{-\frac12\hbar\,q^{j'k}\pdr_{j'}\otimes\nabla_k}f\otimes\psi\big) \\
&=\Del^*\big(e^{-\frac12\hbar\,q^{j'k}\pdr_{j'}\otimes\nabla_k}
 (\pdr_{i'}\otimes1+1\otimes\nabla_{i'}+\mfrac12\hbar\,q^{j'k}
 \pdr_{j'}\otimes\{\nabla_{i'},\nabla_k\})\,f\otimes\psi\big)             \\
&=\Del^*\big(e^{-\frac12\hbar\,q^{j'k}\pdr_{j'}\otimes\nabla_k}
(\pdr_{i'}\otimes1-q^{j'k}q_{i'k}\,\pdr_{j'}\otimes1)\,f\otimes\psi\big)\\
&=0.
\end{align*}
Here we used the curvature \eqref{eqn:curvl} for the third equality.

In contrast, the prequantum action \eqref{eqn:preqtm} does not share this
property.
We illustrate this important difference by an example.
Suppose the fermionic phase space is given by a Euclidean space $(V,q)$ of
dimension $4$ (i.e., $n=2$).
We pick a polarisation $P$ determined by a compatible complex structure and
we choose a complex linear basis so that the non-zero components of $q$ are
$q_{1\bar1}=q_{2\bar2}=\frac12$.
Then the Hilbert space $\cH_P$ consists of sections of the form
$\phi(\tht^1,\tht^2)\exp\frac1{2\hbar}(\bar\tht^1\tht^1+\bar\tht^2\tht^2)$,
where $\phi$ depends on the holomorphic fermionic coordinates $\tht^1,\tht^2$
only.
We take a function $f=\bar\tht^2\bar\tht^1$ and a state
$\psi=\tht^1\tht^2\exp\frac1{2\hbar}(\bar\tht^1\tht^1+\bar\tht^2\tht^2)$.
Then the star product
\[ f*_P\psi=\hbar^2\exp\tfrac1{2\hbar}(\bar\tht^1\tht^1+\bar\tht^2\tht^2) \]
remains a state in $\cH_P$.
However the prequantum action
\[ \hat f(\psi)=(-\bar\tht^1\tht^1\bar\tht^2\tht^2+\hbar\,\bar\tht^1\tht^1+
\hbar\,\bar\tht^2\tht^2)\exp\tfrac1{2\hbar}(\bar\tht^1\tht^1+\bar\tht^2\tht^2) 
\]
is outside $\cH_P$.
A similar problem exists when quantising a function whose Hamiltonian flow
does not preserve the polarisation.
Thus the star product is more preferable than the prequantum action.

Recall now that the bundle $\cO$ over $\cP\subset\ext2V_\bC$ with fibre
$\extb V_\bC^*$ has a flat (but non-product) connection $\nabla^\cO$.
In addition, the bundle $\cH$ over the same base space $\cP$ has a
projectively flat connection $\nabla^\cH$.
We claim that the polarised star products $*_P$ between functions and of
a function on a polarised section are compatible with the connections on
$\cO$ and $\cH$.
In terms of parallel transports $\U^\cO_{P_1,P_0}$ and $\U^\cH_{\{P_t\}}$ in
the bundles $\cO$ and $\cH$ along a smooth path $\{P_t\}_{0\le t\le1}$ in
$\cP$, this means
\begin{equation}\label{eqn:f*psiU}
\U^\cH_{\{P_t\}}(f*_{P_0}\psi)=(\U^\cO_{K_1,K_0}f)*_{P_1}(\U^\cH_{\{P_t\}}\psi)   
\end{equation}
for all $f\in\extb V_\bC^*$, $\psi\in\cH_P$.
Thanks to the differential equations that parallel transports in $\cO$ and
$\cH$ satisfy, it suffices to verify its infinitesimal version
\begin{equation}\label{eqn:infni}
f\,(*_{P+\del P}-*_P)\,\psi=\del(f*_P\psi)-(\del f)*_P\psi-f*_P(\del\psi).
\end{equation}
Indeed, because $\nabla_i$ depends on $t$ through $P_t$, the left hand side of
of \eqref{eqn:infni}, to the first order, is
\begin{align*}
&\quad-\mfrac14\hbar\,\mint_{\!\!0}^1ds\,\Del^*\big(
 e^{-\frac14\hbar(1-s)\Lam_P(d\otimes\nabla)}\,\del(\Lam_P(d\otimes\nabla))
 \,e^{-\frac14\hbar s\Lam_P(d\otimes\nabla)}\,f\otimes\psi\big)            \\
&=-\mfrac14\hbar\,\mint_{\!\!0}^1ds\,\Del^*\big(
 e^{-\frac14\hbar\,\Lam_P(d\otimes\nabla)}\big(\!-\del\Lam_P(d\otimes\nabla)
 -\mfrac14\hbar s[\Lam_P(d\otimes\nabla),\del P(d\otimes\nabla)]\big)
 \,f\otimes\psi\big)                                                        \\
&=-\mfrac12\hbar\mint_{\!\!0}^1ds\,\Del^*\big(
 e^{-\frac14\hbar\Lam_P(d\otimes\nabla)}(-(\del P)^{ij}\pdr_i\otimes\nabla_j
 -\mfrac14\hbar s[q^{i'j}\pdr_{i'}\otimes\nabla_j,
 (\del P)^{k'l'}\pdr_{k'}\otimes\nabla_{l'}])\,f\otimes\psi\big)            \\
&=-\mfrac12\hbar\,\Del^*\big(e^{-\frac14\hbar\,\Lam_P(d\otimes\nabla)}
 (-(\del P)^{ij}\pdr_i\otimes\nabla_j
 -\mfrac12(\del P)^{i'j'}\pdr_{i'}\pdr_{j'}\otimes1)\,f\otimes\psi\big).
\end{align*}
Using the connections $A^\cO$ and $A^\cH$, the right hand side of
\eqref{eqn:infni} is
\begin{align*}
&\quad-\mfrac14\hbar\,(\del P)^{ij}\nabla_i\nabla_j(f*_P\psi)
 +\mfrac14\hbar\,((\del P)^{ij}\pdr_i\pdr_jf
  +(\del P)^{i'j'}\pdr_{i'}\pdr_{j'}f)*_K\psi
  +\mfrac14\hbar f*_P((\del P)^{ij}\nabla_i\nabla_j\psi)                  \\
&=\mfrac14\hbar\,\big(\!-\nabla_i\nabla_j\,
 \Del^*\,e^{-\frac14\hbar\,\Lam_P(d\otimes\nabla)}
 +\Del^*\,e^{-\frac14\hbar\,\Lam_P(d\otimes\nabla)}
 ((\del P)^{ij}(\pdr_i\pdr_j\otimes1+1\otimes\nabla_i\nabla_j)
 +(\del P)^{i'j'}\pdr_{i'}\pdr_{j'}\otimes1)\,\big)f\otimes\psi.
\end{align*}
By the Leibniz rule
$\nabla_i\circ\Del^*=\Del^*\circ(\pdr_i\otimes1+1\otimes\nabla_i)$,
the two sides agree.

The star product is equivariant under the $\SO(V,q)$-actions on functions
and sections.
For any $\gam\in\SO(V,q)$ and $f\in\extb V_\bC^*$, $\psi\in\Gam(\PV,\ell)$,
we have $\gam^\cH(f*_P\psi)=\gam^\cO(f)*_{\gam P\gam^{-1}}\gam^\cH(\psi)$
as an equality in $\Gam(\PV,\ell)$.
If furthermore $\psi\in\cH_P$, then the above equality holds in
$\cH_{\gam P\gam^{-1}}$.
The formula \eqref{eqn:f*psiU} is compatible with the $\SO(V,q)$ action.

\section{Metaplectic correction for fermions}\label{sec:meta}
In the quantisation of bosons, metaplectic (half-form) correction modifies
the prequantum line bundle on a symplectic manifold by tensoring the square
root of the canonical bundle of a polarisation.
It is possible that neither of the two bundles exists as honest bundles, but
their tensor product does, and thus the integrality condition on the symplectic
form is shifted.
If the phase space is a symplectic vector space and the polarisations are
given by compatible linear complex structures, then the bundle of quantum
Hilbert spaces over the space of such polarisations has a projectively flat
connection \cite{ADPW}, while the bundle of metaplectically corrected Hilbert
spaces is flat \cite{Wo,KW}.
For fermions, the bundle of quantum Hilbert spaces over the space of complex
polarisations \cite{Wu15}, and more generally over the space of projections
associated to a pair of transverse Lagrangian subspaces (cf.~\S\ref{sec:qtn}),
is also projectively flat.
But to cancel its curvature, the square root bundle in the metaplectic
correction is the inverse of the one used in the bosonic case \cite{Wu15}.
Curiously, volume element in a fermionic space also transforms in an opposite way.

As before, we assume that the fermionic phase space is given by a Euclidean
space $(V,q)$ of even dimension $2n$ and we denote by $\cP$ the space of
projection operators satisfying \eqref{eqn:projection}.
Over $\cP$ there are tautological vector bundles $\cV$, $\cV'$ of rank $n$
whose fibre over $P\in\cP$ are the complex Lagrangian subspace $\im(P)$,
$\ker(P)$, respectively.
The trivial connection on the product bundle $\cP\times V^\bC$ admitting a
decomposition $\cV\oplus\cV'$ induces a connection $\nabla^\cV$ on $\cV$.
This is a slight generalisation of the setting in \cite{Wu15} where $\cV$
is defined over the subset $\cJ$ of compatible complex structures.
The curvature of $\cV$ is
\[ F^\cV=P\,\del P\wedge\del P\,P,\qquad\mbox{or}\qquad
   (F^\cV)^i_{\;j}=(\del P)^i_{\;k'}\wedge(\del P)^{k'}_{\;j}  \]
in components.
The line bundle $\det(\cV)=\scalebox{1.1}[1.1]{$\mathsf{\Lambda}$}^n\cV$ has
an induced connection $\nabla^{\det(\cV)}$, whose curvature is
\[ F^{\det\cV}=\tr\,(F^\cV)=\tr\,(P\,\del P\wedge\del P\,P)
   =(\del P)^i_{\;j'}\wedge(\del P)^{j'}_{\;i}.  \]
When restricted to $\cJ$, we recover the results in \cite{Wu15}.
The curvatures of $\cV$ and $\det(\cV)$ are, respectively,
\[  F^\cV=-\mfrac14\,P\,\del J\wedge\del J\,P,\qquad
    F^{\det(\cV)}=-\mfrac14\tr\,(P\,\del J\wedge\del J\,P)=\upom_q/\ii,  \]
where $\upom_q$ is the standard K\"ahler form \eqref{eqn:upom} on $\cJ$.

We argue that there is a unique line bundle $(\det\cV)^{1/2}$ over $P$,
called the square root bundle of $\det\cV$, such that
$(\det\cV)^{1/2}\otimes(\det\cV)^{1/2}\cong\det\cV$ \cite{Wu15}.
As $\cP$ has a deformation retract to $\cJ$, it suffices to consider the
bundles over $\cJ$.
Since $c_1(\det\cV)$ is represented by $\upom_q/2\pi$, whose integration on
the generator of $H_2(\cJ)\cong\bZ$ is $2$, the square root bundle of
$\det\cV$ exists.
The uniqueness (up to bundle isomorphisms) is because $\cJ$ is connected
and simply connected.
Furthermore, there is a unique connection on $(\det\cV)^{1/2}$ whose
curvature is $\frac12F^{\det\cV}$.

With metaplectic correction, the quantum Hilbert space in the polarisation
$P\in\cP$ is $\hat\cH_P:=\cH_P\otimes\det(\cV)^{1/2}_P$.
We can define a star product (still denoted by $*_P$) of a function
$f\in\extb V_\bC^*$ on a state $\hat\psi\in\hat\cH_P$
by $f*_P\hat\psi:=(f*_P\psi)\otimes\sqrt\upsilon$, if
$\hat\psi=\psi\otimes\sqrt\upsilon$, $\psi\in\cH_P$,
$\sqrt\upsilon\in(\det\cV)^{1/2}_P$.
We still have associativity $f*_P(g*_P\hat\psi)=(f*_Pg)*_P\hat\psi$ for all
functions $f,g$ on $\PV$ and $\hat\psi\in\hat\cH_P$.
Moreover, $f*_P\hat\psi\in\hat\cH_P$ for all functions $f$ on $\PV$ and
states $\hat\psi\in\hat\cH_P$.
The metaplectically corrected Hilbert spaces $\hat\cH_P$ form a bundle
$\hat\cH:=\cH\otimes(\det\cV)^{1/2}$ over $\cP$.
The curvatures $\cH$ and $(\det\cV)^{1/2}$ cancel, and thus the connection
on $\hat\cH$ is flat.
Moreover, $\hat\cH$ is naturally a global product bundle (see \cite{Wu15}
for the restriction to $\cJ$).
The product structure is reflected in the properties of the metaplectically
corrected intertwining operators.
Along a path $\{P_t\}_{0\le t\le1}$ in $\cP$, the parallel transport
$\U^\cH_{\{P_t\}}$ is a unitary isomorphism from $\cH_{P_0}$ to $\cH_{P_1}$.
Choosing a different path in $\cP$ from $P_0$ to $P_1$, the intertwining
operator $\U^\cH_{\{P_t\}}$ changes by a phase, which is canceled by the
opposite phase from $(\det\cV)^{1/2}$.
Therefore the identification
$\U^{\hat\cH}_{P_1,P_0}\colon\hat\cH_{P_0}\to\hat\cH_{P_1}$ depends only on
the endpoints $P_0,P_1$.
Furthermore, for all $f\in\extb V_\bC^*$, $\hat\psi\in\hat\cH_{P_0}$, we have
\[ (\U^\cO_{P_1,P_0}f)*_{P_1}(\U^{\hat\cH}_{P_1,P_0})
   =\U^{\hat\cH}_{P_1,P_0}(f*_{P_0}\hat\psi).   \]

The action of $\SO(V,\om)$ on $\cP$ lifts to the bundles $\cV$, $\det(\cV)$
preserving the connections $\nabla^\cV$, $\nabla^{\det(\cV)}$ and their
curvatures.
In particular, the $2$-form $F^{\det(\cV)}$ on $\cP$ or $\upom_q$ on $\cJ$
is invariant under $\SO(V,\om)$.
However, only the double cover of $\SO(V,\om)$, the spin group $\Spin(V,\om)$,
acts on $(\det\cV)^{1/2}$ and hence on the metaplectically corrected Hilbert
space bundle $\hat\cH$.
Let $\tilde\gam^{\hat\cH}$ be the action on $\hat\cH$ of a group element
$\tilde\gam\in\Spin(V,q)$ which projects to $\gam\in\SO(V,q)$.
The lifted action preserves the flat connection on $\hat\cH$ and in fact the
product structure of $\hat\cH$ as well.
Moreover, we have
$\tilde\gam^{\hat\cH}\colon\hat\cH_P\to\hat\cH_{\gam P\gam^{-1}}$ for all
$P\in\cP$ and
\[ \tilde\gam^{\hat\cH}\circ\U^{\hat\cH}_{P_1,P_0}=\U^{\hat\cH}
   _{\gam P_1\gam^{-1},\gam P_0\gam^{-1}}\circ\tilde\gam^{\hat\cH} \]
for all $P_0,P_1\in\cP$.
For any $P\in\cP$, $\hat\rho_P(\tilde\gam):=\U^{\hat\cH}_{P,\gam P\gam^{-1}}
\circ\tilde\gam^{\hat\cH}$ defines an honest representation of $\Spin(V,q)$
on $\hat\cH_P$, i.e., for $\tilde\gam,\tilde\gam'\in\Spin(V,q)$, the
equality $\hat\rho_P(\tilde\gam\tilde\gam')=\hat\rho_P(\tilde\gam)\circ
\hat\rho_P(\tilde\gam')$ holds without phase ambiguity.
Finally, we have
\[ \gam^\cO(f)*_{\gam P\gam^{-1}}\tilde\gam^{\hat\cH}(\hat\psi)
   =\tilde\gam^{\hat\cH}(f*_P\hat\psi) \]
for all $f\in\extb V^*_\bC$ and $\hat\psi\in\hat\cH_P$.
All properties remain valid when the bundles on $\cP$ are restricted to $\cJ$
(cf.~\cite{Wu15}).

\setcounter{section}{0}
\renewcommand\thesection{\Alph{section}}
\renewcommand\thesubsection{\thesection.\arabic{subsection}}

\section{Appendix}
\subsection{The Clifford algebra and quantisation maps}\label{sec:ClQ}
The fermionic counterpart of the Weyl algebra is the Clifford algebra of the
fermionic phase space.
Given a Euclidean vector space $(V,q)$, the Clifford algebra $\Cl$ is the
tensor algebra on $V^*_\bC$ modulo the two-sided ideal generated by
$a\otimes b+b\otimes a-\hf\hbar\,q^\sharp(a,b)$, $a,b\in V^*_\bC$, where
again $\hbar>0$ is a fixed number.
Unlike the $\bZ$-graded tensor, symmetric or exterior algebra, the Clifford
algebra is only $\bZ_2$-graded.
So the degree  $|x|$ is defined modulo $2$ but $(-1)^{|x|}$ is well defined
for a homogeneous element $x\in\Cl$.
The super-commutator of $x,y\in\Cl$ is $[x,y]=xy-(-1)^{|x||y|}yx$.
(We omit the multiplication between two elements.)
For example, if $a,b\in V^*_\bC$ are regarded as elements of $\Cl$,
their super-commutator is the anti-commutator $\{a,b\}=ab+ba$ and is equal
to $\hf\hbar\,q^\sharp(a,b)$.

A super-derivation on the Clifford algebra is a linear transformation $D$ on
$\Cl$ such that $D(xy)=(Dx)y+(-1)^{|x|}x(Dy)$ for all $x,y\in\Cl$.
For any $z\in\Cl$, $\ad_z$ defined by $\ad_z(x):=[z,x]$ is an inner
super-derivation.
Another example of super-derivation is by extending the derivative $\pdr_v$
(where $v\in V_\bC$) on $V^*_\bC$, $\pdr_va=\bra a,v\ket$ for $a\in V^*_\bC$,
uniquely to $\Cl$ (still denoted by $\pdr_v$).
In fact $\pdr_v$ is inner, and $\pdr_v=\frac2\hbar\,\ad_{\iota_vq}$, where
$\iota_vq\in V^*_\bC$ is the contraction of $q$ by $v$.
We can check easily $\pdr_va=\frac2\hbar\,\{\iota_vq,a\}$ for $a\in V^*_\bC$;
then by the uniqueness of extension, $\pdr_vx=\frac2\hbar[\iota_vq,x]$
for all $x\in\Cl$.
Equivalently, $\ad_a=\frac\hbar2\,\pdr_{\iota_aq^\sharp}$ or
$[a,x]=\frac\hbar2\,\pdr_{\iota_aq^\sharp}\,x$ for all $x\in\Cl$, where
$\iota_aq^\sharp=a_\mu q^\mn e_\nu\in V_\bC$ is also a contraction.
If $V$ has a linear basis $\{e_\mu\}_{1\le\mu\le m}$, we denote
$\pdm=\pdr_{e_\mu}$ as before.
Then $\ad_a=\frac\hbar2\,q^\mn a_\mu\pdn$ on $\Cl$ for all $a\in V^*_\bC$.

We define a map $\varrho_0\colon V^*_\bC\to\End(\Cl)$ by
$\varrho_0(a)x=\hf(ax+(-1)^{|x|}xa)$ for all $a\in V^*_\bC$, $x\in\Cl$.
The equality
\begin{equation}\label{eqn:varrho1}
\pdr_v\circ\varrho_0(a)+\varrho_0(a)\circ\pdr_v=\bra a,v\ket\id
\end{equation}
of operators on $\Cl$ can be easily verified for
$v\in V_\bC$ and $a\in V^*_\bC$.
Since $\{\varrho_0(a),\varrho_0(b)\}=0$ for all $a,b\in V^*_\bC$, the map
extends uniquely to (using the same notation)
$\varrho_0\colon\extb V^*_\bC\to\End(\Cl)$ such that
$\varrho_0(1)$ is the identity operator and $\varrho_0(f\wedge g)=
\varrho_0(f)\circ\varrho_0(g)=(-1)^{|f||g|}\varrho_0(g)\circ\varrho_0(f)$
for all $f,g\in\extb V^*_\bC$.
We claim that
\begin{equation}\label{eqn:varrho}
\pdr_v\circ\varrho_0(f)-(-1)^{|f|}\varrho_0(f)\circ\pdr_v=\varrho_0(\pdr_vf)
\end{equation}
for all $v\in V_\bC$ and $f\in\extb V^*_\bC$;
the left hand side is the super-commutator $[\pdr_v,\varrho_0(f)]$ of linear
transformations on the graded vector space $\Cl$.
The identity \eqref{eqn:varrho} can be proved by induction on $|f|$.
It holds obviously if $|f|=0$ or $f=1$ and it is \eqref{eqn:varrho1}
if $|f|=1$ or $f=a$ for some $a\in V^*_\bC$.
If $f,g\in\extb V^*_\bC$ satisfy \eqref{eqn:varrho}, then so does
$f\wedge g$ because
\begin{align*}
&[\pdr_v,\varrho_0(f\wedge g)]=[\pdr_v,\varrho_0(f)\circ\varrho_0(g)]
=[\pdr_v,\varrho_0(f)]\circ\varrho_0(g)+(-1)^{|f|}\varrho_0(f)\circ
 [\pdr_v,\varrho_0(g)]                                              \\
=&\varrho_0(\pdr_vf)\circ\varrho_0(g)+(-1)^{|f|}\varrho_0(f)\circ
 \varrho_0(\pdr_vg)
=\varrho_0(\pdr_vf\wedge g+(-1)^{|f|}f\wedge\pdr_vg)
=\varrho_0(\pdr_v(f\wedge g)).
\end{align*}

The quantisation map $\Q_0\colon\extb V^*_\bC\to\Cl$ sends a fermionic
function $f\in\extb V^*_\bC$ to $\Q_0(f):=\varrho_0(f)1$ in the Clifford
algebra $\Cl$.
This association of an element $\Q_0(f)$ in the quantum operator algebra to a
classical function $f$ is the counterpart of the bosonic Weyl ordering since
for all $a\in V^*_\bC$, and $x\in\Cl$, $\varrho_0(a)x=\hf(ax+(-1)^{|x|}xa)$ is
the average of the two orderings of $a,x$ in the graded sense.
The two spaces $\extb V^*_\bC$ and $\Cl$ are of the same dimension and $\Q_0$
is a linear isomorphism whose inverse is the standard symbol map from the
Clifford algebra to the exterior algebra.
By applying the identity \eqref{eqn:varrho} to $1\in\Cl$ and using $\pdr_v1=0$,
we obtain,
\begin{equation}\label{eqn:pdrQ0}
\pdr_v\Q_0(f)=\pdr_v(\varrho_0(f)1)=\varrho_0(\pdr_vf)1=\Q_0(\pdr_vf)
\end{equation}
for all $v\in V_\bC$ and $f\in\extb V^*_\bC$.
Equivalently, if $a\in V^*_\bC$, we have
\[  \ad_a\!\Q_0(f)=\mfrac\hbar2\,\Q_0(\pdr_{\iota_aq^\sharp}f)
=\mfrac\hbar2\,q^\mn a_\mu\Q_0(\pdn f).  \]

The quantisation map $\Q_0$ intertwines the fermionic Moyal product
\eqref{eqn:moyal} and the Clifford product.
That is,
\begin{equation}\label{eqn:Q0*0}
\Q_0(f)\,\Q_0(g)=\Q_0(f*_0g)
\end{equation}
for all $f,g\in\extb V^*_\bC$.
Clearly, it holds for all $g$ if $f=1$.
Suppose \eqref{eqn:Q0*0} is true for some $f$, then for any $a\in V^*_\bC$,
\begin{align*}
&\Q_0(a\wedge f)\,\Q_0(g)=(\varrho_0(a)\Q_0(f))\Q_0(g)
=\varrho_0(a)(\Q_0(f)\Q_0(g))+(-1)^{|f|}\hf\,\Q_0(f)\ad_a\Q_0(g)       \\
=&\varrho_0(a)\Q_0(f*_0g)+(-1)^{|f|}\tfrac\hbar4\,q^\mn a_\mu\Q_0(f)
 \Q_0(\pdn g)
=\Q_0\big(a\wedge(f*_0g)+(-1)^{|f|}\tfrac\hbar4\,q^\mn a_\mu f*_0(\pdn g)
 \big),
\end{align*}
which equals $\Q_0((a\wedge f)*_0g)$ by \eqref{eqn:awedgef}.
Thus we have verified \eqref{eqn:Q0*0} by induction on $|f|$.
Associativity of the fermionic Moyal product $*_0$ is now a consequence
of \eqref{eqn:Q0*0}.

To obtain quantisation map for the star product $*_K$ in \eqref{eqn:stark},
we note that if $K,K'\in\ext2V_\bC$, the operator
$\U^\cO_{K',K}=e^{-\frac\hbar8(K'-K)^\mn\pdm\pdn}=(\U^\cO_{K,K'})^{-1}$ acts
on the Clifford algebra $\Cl$ since $\pdm$ acts on it as a super-derivation.
For any $K\in\ext2V_\bC$, we define $\varrho_K\colon V^*_\bC\to\End(\Cl)$ by
$\varrho_K(a):=\U^\cO_{0,K}\circ\varrho_0(a)\circ\U^\cO_{K,0}
=\varrho_0(a)+\frac\hbar8\,K^\mn a_\mu\pdn$, where $a\in V^*_\bC$.
Just like $\varrho_0$, the map $\varrho_K$ satisfies
$\{\varrho_K(a),\varrho_K(b)\}=0$ for all $a,b\in V^*$, and hence
it extends to $\varrho_K\colon\extb V^*_\bC\to\End(\Cl)$ such that
$\varrho_K(f\wedge g)=\varrho_K(f)\circ\varrho_K(g)$ for all $f,g$.
In fact, $\varrho_K(f)=\U^\cO_{0,K}\circ\varrho_0(f)\circ\U^\cO_{K,0}$ if
$f\in\extb V^*_\bC$.
The quantisation map $\Q_K\colon\extb V_\bC^*\to\Cl$ for a general
$K\in\ext2V$ is $\Q_K(f):=\varrho_K(f)1$.
Using \eqref{eqn:pdrQ0}, we have
$\Q_K(f)=\U^\cO_{0,K}\Q_0(f)=\Q_0(\U^\cO_{0,K}f)$, and thus for all
$a\in V_\bC^*$, $\Q_K(a)=\Q_0(a)$.
Furthermore, we have
\[ \Q_K(f)\,\Q_K(g)=\Q_0(\U^\cO_{0,K}f)\,\Q_0(\U^\cO_{0,K}g)
   =\Q_0((\U^\cO_{0,K}f)*_0(\U^\cO_{0,K}g))
   =\Q_0(\U^\cO_{0,K}(f*_Kg))=\Q_K(f*_Kg)             \]
for all $f,g\in\extb V^*_\bC$.
Consequently, $[\Q_K(f),\Q_K(g)]=\hbar\,\Q_K(\{f,g\})+o(\hbar^2)$ for all
$K\in\ext2 V_\bC$.

If $\{K_t\}_{0\le t\le1}$ is a smooth path in $\ext2V_\bC$ from $K_0$ to $K_1$,
then $\Q_{K_t}(\U^\cO_{K_t,K_0}f)$ is constant in $\Cl$.
In particular, $\Q_{K_1}(\U^\cO_{K_1,K_0}f)=\Q_{K_0}(f)$ for all
$f\in\extb V^*_\bC$.
Consider a map $\cO\to\ext2V_\bC\times\Cl$ of bundles over $\ext2V_\bC$ by
applying $\Q_K$ to the fibre over $K$.
The above property means that the flat connection $\nabla^\cO$ is mapped to
the trivial connection on the product $\Cl$-bundle over $\ext2V_\bC$.

The special orthogonal group $\SO(V,q)$ acts as (inner) automorphisms on the
Clifford algebra $\Cl$; the action $\gam^\C$ of $\gam\in\SO(V,q)$ on $\Cl$ is
uniquely determined by $\gam^\C(1)=1$ and $\gam^\C(a)=a\circ\gam^{-1}$ for all
$a\in V_\bC^*$.
The maps $\varrho_0$ and $\Q_0$ constructed above are equivariant under
$\SO(V,q)$.
That is, for all $\gam\in\SO(V,q)$, $f\in\extb V^*_\bC$, we have
$\varrho_0(\gam^\cO(f))=\gam^\C\circ\varrho_0(f)\circ(\gam^\C)^{-1}$
and $\gam^\C(\Q_0(f))=\Q_0(\gam^\cO(f))$.
More generally, if $K\in\ext2V_\bC$, we have $\varrho_{(\gam\otimes\gam)K}
(\gam^\cO(f))=\gam^\C\circ\varrho_K(f)\circ(\gam^\C)^{-1}$ and
$\gam^\C(\Q_K(f))=\Q_{(\gam\otimes\gam)K}(\gam^\cO(f))$.
The lifted action of $\SO(V,q)$ on $\cO$ and $\ext2V_\bC\times\Cl$ are
compatible with the bundle map between them.

\subsection{The fermionic Stratonovich-Weyl quantiser and star products}
In bosonic theory, the quantisation map can be expressed as an integral over
the phase space of an operator valued function called the Stratonovich-Weyl
kernel or quantiser \cite{We,St}.
For a fermionic system whose phase space is given by a Euclidean space $(V,q)$
of dimension $m$, the analogue of the Stratonovich-Weyl quantiser \cite{GGPT}
is a $\Cl$-valued fermionic function $\Om_0(\tht)$ on $\PV$, or an element
$\Om_0\in\Cl\,\hat\otimes\,\extb V^*_\bC$.
Let $\tht^\mu$ ($\mu=1,\dots,m$) be the fermionic coordinates on $\PV$ under
a linear babis $\{e_\mu\}$ of $V$ such that $q_\mn=\del_\mn$ and
$\eps^{12\cdots m}=1$ and let $\hat\tht^\mu:=Q_0(\tht^\mu)$ be the
quantisation of $\tht^\mu$.
Then $\Om_0$ can be written as
$\Om_0(\tht)=(\tht^1-\hat\tht^1)\cdots(\tht^m-\hat\tht^m)$.
The quantisation map is
\begin{equation}\label{eqn:qSW}
\Q_0(f)=\mint_{\!\!\PV}\,d\tht\,\Om_0(\tht)f(\tht)
\end{equation}
for all fermionic functions $f\in\extb V^*_\bC$.
In particular,
$\Q_0(\tht^{\mu_1}\cdots\tht^{\mu_p})=\hat\tht^{\mu_1}\cdots\hat\tht^{\mu_p}$
if $\mu_1,\dots,\mu_p$ are mutually distinct.

We continue to assume that the dimension $m=2n$ of $V$ is even.
A supertrace on the Clifford algebra $\Cl$ is determined up to a scalar factor
by requiring its vanishing on the graded commutators.
The factor can be fixed by the supertrace in the spinor representation, the
quantum Hilbert space of the fermionic system (cf.~\S\ref{sec:qtn}).
In the above fermionic coordinates, we have
\[  \str(\hat\tht^{\mu_1}\hat\tht^{\mu_2}\cdots\hat\tht^{\mu_p})=0\quad(p<m),
\qquad\str(\hat\tht^1\hat\tht^2\cdots\hat\tht^{2n})=(\ii\hbar/2)^n. \]
It follows that
\[ \str(\Q_0(f))
 =\big(\mfrac{\ii\hbar}2\big)^{\!n}\mint_{\!\!\PV}\,d\tht\,f(\tht)  \]
for all fermionic functions $f$ on $\PV$.
In addition, we have the identity
\[ \str(\Om_0(\tht)\Om_0(\tht'))
   =\big(\mfrac{\ii\hbar}2\big)^{\!n}\del(\tht-\tht'),  \]
where $\del(\tht-\tht')=(\tht^1-\tht'{}^1)\cdots(\tht^{2n}-\tht'^{2n})$ is
the fermionic delta-function satisfying
\[  \mint_{\!\!\Pi\bR}\,d\tht'\,\del(\tht-\tht')f(\tht')=f(\tht). \]
As in the bosonic theory \cite{AW}, the Stratonovich-Weyl quantiser $\Om_0$
is a quantised version of the delta-function.
The inverse of the quantisation map can be expressed as a supertrace: for any
$a\in\Cl$, we have \cite{GGPT}
\[  \Q_0^{-1}(a)=\big(\mfrac2{\ii\hbar}\big)^{\!n}\str(\Om_0(\tht)a). \]

The Moyal product of two fermionic functions $f$ and $g$ can be written as
\[  (f*_0g)(\tht)
   =\big(\mfrac2{\ii\hbar}\big)^{\!n}\str(\Q_0(f*_0g)\Om_0(\tht))
   =\big(\mfrac2{\ii\hbar}\big)^{\!n}\str(\Q_0(f)\Q_0(g)\Om_0(\tht))  \]
using the fermionic Stratonovich-Weyl quantiser.
By \eqref{eqn:qSW} and the identity
\[ \str(\Om_0(\tht')\Om_0(\tht'')\Om_0(\tht))=\mfrac{(\ii\hbar)^{3n}}{2^{5n}}
\exp\big[\tfrac4\hbar(q(\tht,\tht')+q(\tht',\tht'')+q(\tht'',\tht))\big], \]
the Moyal product can be represented by a Berezin integral \cite{GGPT}
\begin{align*}
(f*_0g)(\tht)&=\big(\mfrac{\ii\hbar}4\big)^{2n}\mint_{\!\!\PV\times\PV}
  \,d\tht'd\tht''f(\tht')g(\tht'')\exp\big[\tfrac4\hbar(q(\tht,\tht')
  +q(\tht',\tht'')+q(\tht'',\tht))\big]  \\
&=(-1)^n\mint_{\!\!\PV\times\PV}\,d\tht'd\tht''f\big(\tht+\tfrac{\sqrt\hbar}
  2\tht'\big)g\big(\tht+\tfrac{\sqrt\hbar}2\tht''\big)\exp q(\tht',\tht''),
\end{align*}
which is equivalent to \eqref{eqn:moyal}.

For an antisymmetric bivector $K\in\ext2V_\bC$, the quantisation map is
\begin{equation}\label{eqn:qSWK}
\Q_K(f)=\mint_{\!\!\PV}\,d\tht\,\Om_0(\tht)(\U_{0,K}^\cO f)(\tht)
=\mint_{\!\!\PV}\,d\tht\,\Om_K(\tht)f(\tht),
\end{equation}
where $\Om_K(\tht):=e^{\frac\hbar8K^\mn\pdr_\mu\pdr_\nu}\Om_0(\tht)$.
Here $\pdr_\mu=\pdr/\pdr\tht^\mu$, and with the fermionic coordinates
$\tht'{}^\mu$ and $\tht''{}^\mu$, we write $\pdr'_\mu=\pdr/\pdr\tht'{}^\mu$
and $\pdr''_\mu=\pdr/\pdr\tht''{}^\mu$.
Using the identity
\[ \str(\Om_{-K}(\tht)\Om_K(\tht'))=\big(\mfrac{\ii\hbar}2\big)^{\!n}
    e^{\frac\hbar8K^\mn(\pdr'_\mu\pdr'_\nu-\pdr_\mu\pdr_\nu)}\del(\tht-\tht')
   =\big(\mfrac{\ii\hbar}2\big)^{\!n}\del(\tht-\tht'),  \]
we obtain the inverse of the quantisation map $\Q_K$ for $a\in\Cl$,
\[ \Q_K^{-1}(a)=\big(\mfrac2{\ii\hbar}\big)^{\!n}\str(\Om_{-K}(\tht)a). \]

The star product $*_K$ of two fermionic functions $f,g$ on $\PV$ is
\[ (f*_Kg)(\tht)
   =\big(\mfrac2{\ii\hbar}\big)^{\!n}\str(\Q_K(f*_0g)\Om_{-K}(\tht))
   =\big(\mfrac2{\ii\hbar}\big)^{\!n}\str(\Q_K(f)\Q_K(g)\Om_{-K}(\tht)).  \]
Using \eqref{eqn:qSWK} and the identity
\begin{align*}
\str&(\Om_K(\tht')\Om_K(\tht'')\Om_{-K}(\tht))=e^{\frac\hbar8K^\mn
   (\pdr'_\mu\pdr'_\nu+\pdr''_\mu\pdr''_\nu-\pdr_\mu\pdr_\nu)}
   \str(\Om_0(\tht')\Om_0(\tht'')\Om_0(\tht))                      \\
&=\mfrac{(\ii\hbar)^{3n}}{2^{5n}}\exp\big[\tfrac4\hbar
 (q(\tht,\tht')+q(\tht',\tht'')+q(\tht'',\tht)+\hf K^\flat(\tht'-\tht)
 +\hf K^\flat(\tht''-\tht)-\hf K^\flat(\tht'-\tht''))\big],
\end{align*}
where $K^\flat\in\medext V_\bC^*$ is the antisymmetric bilinear form on
$V_\bC$ induced by $K$ and $q$ and $K^\flat(\tht)=K^\flat(\tht,\tht)$,
we obtain
\begin{align*}
(f*_Kg)(\tht)&=\big(\mfrac2{\ii\hbar}\big)^{\!n}\mint_{\!\!\PV\times\PV}\,
d\tht'd\tht''f(\tht')g(\tht'')\str(\Om_K(\tht')\Om_K(\tht'')\Om_{-K}(\tht))\\
&=(-1)^n\mint_{\!\!\PV\times\PV}\,d\tht'd\tht''
  f\big(\tht+\tfrac{\sqrt\hbar}2\tht'\big)
  g\big(\tht+\tfrac{\sqrt\hbar}2\tht''\big)
  \exp\big[(q+K^\flat)(\tht',\tht'')\big],
\end{align*}
expressing the star product \eqref{eqn:stark} as a fermionic integral.
Similar formulae exist for bosons with a symmetric $K$.

For $\gam\in\SO(V,q)$, we have $(\gam^\C\hat\otimes\gam^\cO)\Om_0=\Om_0$
and, more generally,
$(\gam^\C\hat\otimes\gam^\cO)\Om_K=\Om_{(\gam\otimes\gam)K}$ for all
$K\in\ext2V_\bC$.
This is consistent with the properties of the quantisation maps $\Q_0$,
$\Q_K$ described in \S\ref{sec:ClQ}.

\medskip\noindent
{\em Acknowledgments.}
The author is supported in part by a grant 106-2115-M-007-005-MY2 from MOST
(Taiwan).
He thanks M.\;Pflaum, M.\;Schlichenmaier and A.\;Yoshioka for discussions on
star products in the bosonic case.

\end{document}